\documentclass[fleqn]{article}
\usepackage{fleqn}
\usepackage{graphics}
\usepackage{epsfig}
\usepackage{amsmath}
\usepackage{amssymb}
\usepackage{calc}
\newcommand{\beq}{\begin{equation}}
\newcommand{\eeq}{\end{equation}}
\newcommand{\bea}{\begin{eqnarray}}
\newcommand{\eea}{\end{eqnarray}}
\begin{document}
\Large
\begin{center}
{\bf  The Coupled Cluster Method and Entanglement}
\end{center}
\large
\vspace*{-.1cm}
\begin{center}
 P\'eter L\'evay$^{1}$ , Szilvia Nagy$^{2}$, J\'anos Pipek$^{3}$ and G\'abor S\'arosi$^{3}$
 \end{center}
 \vspace*{-.4cm} \normalsize
 \begin{center}

 $^{1}$Department of Theoretical Physics, Institute of Physics, Budapest University of\\
 Technology and Economics and MTA-BME Condensed Matter Research Group, H-1521 Budapest, Hungary

$^2$Sz\'echenyi Istv\'an University, Department of Telecommunications, H9026 Gy\H or Egyetem t\'er 1

$^{3}$Department of Theoretical Physics, Institute of Physics, Budapest University of\\
 Technology and Economics, H-1521 Budapest, Hungary

\vspace*{.0cm}

\vspace*{.2cm} (19. July 2016)

\end{center}

\vspace*{-.3cm} \noindent \hrulefill

\vspace*{.1cm} \noindent {\bf Abstract:}
The Coupled Cluster (CC) and full CI expansions are studied for three fermions with
six  and seven modes. Surprisingly the CC expansion is tailor made
to characterize the usual SLOCC entanglement classes. It means that
the notion of a SLOCC transformation shows up quite naturally as a
one relating the CC and CI expansions, and going from the CI
expansion to the CC one is equivalent to obtaining a
form for the state where the structure of
the entanglement classes is transparent.
In this picture entanglement is characterized by the parameters of the cluster operators
describing transitions from occupied states to singles, doubles and triples of non occupied ones.
Using the CC parametrization of states in the seven mode case we give 
a simple formula for the unique SLOCC invariant $\mathcal{J}$.
Then we consider a perturbation problem featuring a state from the 
unique SLOCC class characterized by $\mathcal{J}\neq 0$.
For this state
with entanglement generated by doubles
we investigate the phenomenon of changing the entanglement type due to the perturbing effect of triples.
We show that there are states with real amplitudes such that their entanglement encoded into configurations of clusters of doubles is protected from errors generated by triples.
Finally we put forward a proposal to use the parameters of the cluster operator describing transitions to doubles
for entanglement characterization.
Compared to the usual SLOCC classes this provides a coarse grained approach to fermionic entanglement.

 \vspace*{.3cm} \noindent
 \hspace*{1.95cm} 

 \vspace*{-.2cm} \noindent \hrulefill

\section{Introduction}

According to the current paradigm quantum entanglement can be
regarded as a new and efficient resource\cite{Nielsen}. Suppose that we have a
multipartite quantum system consisting of a number of subsystems.
Entanglement can then be created between these subsystems via
applying a set of {\it global} operations on the full system by
switching on certain interactions. These can be effected via using
successively a {\it discrete} sequence of quantum gates, or via
applying a {\it continuous} evolution operator generated by some
Hamiltonian. The particular type of entanglement obtained in this
way is {\it defined} by prescribing a certain set of {\it local
operations} acting on the {\it subsystems} that will result in
{\it equivalent} states of the full system.

The most common operations used in quantum theory are of course
the {\it unitary ones} resulting in unitary quantum gates and
evolutions\cite{Nielsen}. These are the ones leaving invariant the scalar
products inherently connected to the structure of the multipartite
space of states. Choosing the prescribed set as the {\it local}
versions of such operations gives rise to the definition of
entanglement classes as the set of equivalence classes of the
space of states of the full system under the action of the group
of {\it local unitary} (LU) transformations. Hence, two states are
possessing the same entanglement if and only if they can be
converted to each other via some local unitary transformation\cite{Plenio}.

However, for practical purposes it turned out that this
classification scheme is too restrictive\cite{Thapli,Dur}. First of all, local
unitary classification is providing a fine graining to the full
set of states, whereas sometimes a coarse graining would be more
desirable. Furthermore in quantum information processing
manipulations of more general types then unitary ones are also
allowed. These are physical manipulations (including communication
between parties applying classical channels) resulting in the
conversion of states back and forth with (generally different)
success probabilities. These operations are called SLOCC
operations (stochastic local operations and classical
communication). They can be represented mathematically via the
{\it local} action of the general linear group of {\it invertible}
transformations of which the set of local unitary ones is merely a
subgroup. The SLOCC entanglement classes are then defined as the
equivalence classes of the set of states of the full system under
the action of the group of local invertible SLOCC transformations.
This new set of equivalence classes gives rise to a coarse
graining of the space of states of the multipartite system.

Notice however, that unlike for a local unitary transformation the
physical meaning for a particular local invertible one is not so
obvious. Let us take the example of a multiqubit system.  A local
unitary transformation on a particular qubit is represented by a
$2\times 2$ unitary matrix which can be implemented by using
physical equipment like Stern-Gerlach magnets, certain
arrangements modelled by $2\times 2$ Hamiltonians like the ones
featuring the interaction of $1/2$ spins with external magnetic
fields etc. On the other hand though being an elementary local
invertible one the non unitary transformation
$\begin{pmatrix}1&a\\0&1\end{pmatrix}, a\in{\mathbb C}$ is not so
easy to implement by a natural physical setup.

The notion of SLOCC classes can also be generalized to systems
with indistinguishable constituents\cite{Schliemann1,Schliemann2}. Since the subsystems in this
case are indistinguishable the local invertible transformations
representing SLOCC manipulations should be identical. One can then
ask what kind of physical representations of SLOCC transformations
are available in this context?
 The aim of the present paper is to
 show that in the special case of choosing our system with
 indistinguishable constituents to be a fermionic one the notion of
 SLOCC transformations is intimately connected to a physically well
 established method used by quantum chemists for a long time. This
 method is the Coupled Cluster (CC) method\cite{CCmethod,CCM2}.

In the CC method the usual expansion of the state vector
$\vert\Psi\rangle$ of the fermionic system given as a linear
combination of Slater determinants (CI expansion) is replaced by a
new one. The CC expansion is constructed via the action of a set
of commuting cluster operators $e^{\hat{T}_i}$ $i=1,2,\dots$ on a
special Slater determinant $\vert\Psi_0\rangle$ singled out by
special physical considerations. The single particle states
comprising $\vert\Psi_0\rangle$ are called the {\it occupied}
ones.
 Then the cluster
 operators $\hat{T}_i$
  describe single, double, triple etc. transitions from
  the occupied single particles states to non-occupied ones. Hence
  in this CC picture the state $\vert\Psi\rangle$ is regarded as a
  deviation from the distinguished state $\vert\Psi_0\rangle$
  effected by multiple transitions from occupied states to non
  occupied ones.

In this paper by employing three-fermion systems with six\cite{LevVran,Chen0,SarLevCKW} and
seven\cite{SarLev2} single particle states (modes) we would like to draw the
readers attention to the fact that SLOCC transformations show up
quite naturally within the framework of the CC method. In this
context invertible transformations such as the aforementioned ones
of upper triangular form are having a clear cut physical meaning.
We also demonstrate that going from the CI expansion to the CC one
plays the role of obtaining a form for a fermionic state
where the structure of the SLOCC entanglement classes is
transparent. The reason for restricting our attention merely to
these special cases stems from the fact that apart from these ones
(and the much more elaborate cases of three and four fermions with
eight modes\cite{Chen1,LevHolw}) the structure of the SLOCC classes and its invariants
and covariants needed for our considerations is not known. Recall
however, that precisely these special cases were the ones where
the Borland-Dennis inequalities\cite{Borland,Russkai} of quantum chemistry have been
established. As is well-known by now these studies have given rise
to recent developments culminating in the introduction of the
generalized Pauli principle in connection with the famous N-representability problem\cite{Klyachko,Kly2} 
and the idea of entanglement polytopes\cite{Christ,pinning}.
 In this spirit we conjecture that though illustrated merely in the simplest multipartite cases our
results should have a generalization valid also for multi-fermionic
systems with an arbitrary number of modes. We hope that our
observations will pave the
way for interesting future applications.

The organization of this paper is as follows. In Section 2. we
consider the case of three-fermionic systems with six modes. First
by giving an explicit dictionary between the corresponding
expansion coefficients we relate the CI and CC expansions. Then we
look at the structure of the unique SLOCC invariant $\mathcal{D}$
of fourth order serving as a convenient measure of entanglement.
We observe that there is a dramatic simplification in the
structure of this invariant in the CC picture. In this case no
contribution from the cluster operators $\hat{T}_1$ appears. It
turns out that the reason for this is the fact that 
$\hat{T}_1$
generates SLOCC transformations, hence the corresponding term $e^{\hat{T}_1}$ can
be removed from the expansion of $\vert\Psi\rangle$. Next we show
that the remaining expansion coefficients related to $\hat{T}_2$
and $\hat{T}_3$ are characterizing the four nontrivial SLOCC
entanglement classes in a simple manner. Section 3. is devoted to
a study of the seven mode generalization of the results of Section
2. After recalling the structure of the basic invariant
$\mathcal{J}$ of seventh order, the  covariants and the SLOCC
classes we again relate the CC and CI expansions. In the CC
picture we find a surprisingly simple expression for
$\mathcal{J}$. As a demonstration how the entanglement in this
case is manifested in the structure of the cluster operators
$\hat{T}_2$ and $\hat{T}_3$ we consider a perturbation problem.
As a first step we choose a maximally entangled state defined by the condition $\mathcal{J}\neq 0$ 
which is generated merely by the
expansion coefficients of $\hat{T_2}$ (doubles). Then we modify this state
by adding also the contributions coming from $\hat{T_3}$ (triples).
This perturbation due to triples is characterized by four complex numbers.
As a next step we derive a constraint for the triples to be able to induce a transition to a different SLOCC class.
It turns out that the constrained set of parameters define a six dimensional complex manifold which is known as the deformed conifold, i.e. a deformation of a six dimensional cone.
Taking the perturbation parameters to be real reveals a new effect. 
If we change the sign of certain parameters of the doubles then
the perturbation due to triples
cannot induce a transition to a different SLOCC class.
Hence in this special case the maximal entanglement encoded into the parameters 
of the doubles is protected from the perturbating effect of the triples.
In section 4. we put forward a proposal to use the parameters of the cluster operators describing transitions to doubles
for entanglement characterization.
Compared to the usual SLOCC classes this provides a coarse grained approach to fermionic entanglement.
Using our model systems familiar from Sections 2. and 3. and the new system of $4$ fermions with $8$ modes we offer some mathematical evidence supporting our proposal.
Our conclusions and some comments are left to Section 5.
For the convenience of the reader we put some of the technical details to an Appendix.

\section{Three fermions with six modes}

\subsection{Relating the CI and CC expansions}

We consider three fermions with six single particle states or modes.
Let $V$ be a six dimensional complex vector space and $V^{\ast}$ its dual.
We regard $V\simeq {\mathbb {C}}^6$ with $\{e_{\mu}\}, \mu=1,2,\dots 6$ the canonical basis and $\{e^{\mu}\}$ is the dual basis. Elements of $V$ will be called {\it single particle states} or modes.
We tacitly assume that $V$ is a Hilbert space also equipped with a Hermitian inner product, but we will not make use of this extra structure.
We also introduce the twelve dimensional vector space
\beq
\mathcal {V}\equiv V\oplus V^{\ast}
\label{nu}.
\eeq
\noindent
An element of $\mathcal{V}$ is of the form $x=v+\alpha$ where $v$ is a vector
and $\alpha$ is a linear form.
In the case of fermions according to the method of second quantization to any element $x\in\mathcal{V}$ one can associate a linear operator $\hat{x}$ acting on the fermionic Fock space $\mathcal{F}$ as follows.
We define fermionic creation and annihilation operators as
\beq
\hat{e}^{\mu}\equiv\hat{p}^{\mu},\qquad \hat{e}_{\mu}\equiv\hat{n}_{\mu},\qquad \mu=1,\dots 6
\label{defioper}
\eeq
\noindent
satisfying the usual
fermionic anticommutation relations \beq
\{\hat{p}^{\mu},\hat{n}_{\nu}\}=\delta^{\mu}_{\nu}\hat{1},\qquad
\{\hat{p}^{\mu},\hat{p}^{\nu}\}=\{\hat{n}_{\mu},\hat{n}_{\nu}\}=0.
\label{anticommute} \eeq \noindent 
We define the vacuum state  $\vert 0\rangle\in\mathcal{F}$ by the property
\beq
\hat{n}_{\mu}\vert 0\rangle =0.
\label{vacuum}
\eeq
\noindent
Then $\mathcal{F}$ is spanned by the basis vectors\
\beq
\hat{p}^{\mu_1}\hat{p}^{\mu_2}\cdots\hat{p}^{\mu_k}\vert 0\rangle,\qquad 1\leq\mu_1<\mu_2<\cdots<\mu_k\leq 6.
\label{fockbasis}
\eeq
\noindent
In particular for $k=3$ an unnormalized three
fermion state with six modes can be written in the form \beq
\vert\psi\rangle=\frac{1}{3!}\psi_{\mu\nu\rho}\hat{p}^{\mu}\hat{p}^{\nu}\hat{p}^{\rho}\vert
0\rangle \label{kifejt} \eeq \noindent where $\psi_{\mu\nu\rho}$
is totally antisymmetric having $20$ independent complex
amplitudes. 

The SLOCC group is the one of invertible $6\times 6$ matrices with complex entries i.e. $GL(6,\mathbb{C})$. It is acting on a
state via transforming its amplitudes as \beq
\psi_{\mu\nu\rho}\mapsto {S_{\mu}}^{\mu^{\prime}}
{S_{\nu}}^{\nu^{\prime}} {S_{\rho}}^{\rho^{\prime}}
\psi_{\mu^{\prime}\nu^{\prime}\rho^{\prime}},\qquad S\in
GL(6,\mathbb{C}). \label{sixtrafo}
 \eeq
\noindent

Let us now split the single particle states to ones that are occupied and not occupied.
The ones that are occupied will be labelled as
\beq
i,j,k=1,2,3
\eeq
\noindent
and the ones that are not as
\beq
a,b,c=\overline{1},\overline{2},\overline{3}.
\eeq
\noindent
Hence we have a new labelling for the six modes as
\beq
\{1,2,3,4,5,6\}\equiv \{1,2,3,\overline{1},\overline{2},\overline{3}\}.
\label{konvencio}
\eeq
\noindent
We take the Hartee-Fock (HF) state as the Slater determinant
    \beq
\vert\psi_0\rangle\equiv \hat{p}^1\hat{p}^2\hat{p}^3\vert
0\rangle. \label{HFstate} \eeq \noindent
We see that this distinguished element of $\mathcal{F}$ is built up
from single particle states that are occupied.
Now
the Coupled Cluster
(CC) and full CI expansions are respectively \beq
\vert\psi\rangle=e^{\hat{T}_1+\hat{T}_2+\hat{T}_3}\vert\psi_0\rangle
\label{ccexp} \eeq \noindent and \beq
\vert\psi\rangle=(\hat{1}+\hat{C}_1+\hat{C}_2+\hat{C}_3)\vert\psi_0\rangle.
\label{fci} \eeq \noindent Here \beq
\hat{T}_1={T_a}^i\hat{p}^a\hat{n}_i, \qquad
\hat{T}_2=\frac{1}{4}{T_{ab}}^{ij}\hat{p}^a\hat{n}_i\hat{p}^b\hat{n}_j
\qquad
\hat{T}_3={T_{\overline{1}\overline{2}\overline{3}}}^{123}\hat{p}^{\overline{1}}\hat{n}_1
\hat{p}^{\overline{2}}\hat{n}_2 \hat{p}^{\overline{3}}\hat{n}_3
\label{tk} \eeq \noindent and

\beq \hat{C}_1={C_a}^i\hat{p}^a\hat{n}_i, \qquad
\hat{C}_2=\frac{1}{4}{C_{ab}}^{ij}\hat{p}^a\hat{n}_i\hat{p}^b\hat{n}_j
\qquad
\hat{C}_3={C_{\overline{1}\overline{2}\overline{3}}}^{123}\hat{p}^{\overline{1}}
\hat{n}_1 \hat{p}^{\overline{2}}\hat{n}_2
\hat{p}^{\overline{3}}\hat{n}_3 \label{tkc} \eeq \noindent
are called the cluster operators. Notice that by construction 
the set of operators $\{\hat{T}_1,\hat{T}_2,\hat{T}_3\}$ called {\it singles}, {\it doubles} and {\it triples} are commuting.

For the expansion coefficients of the "T"s and "C"s it will be
useful to adopt the following labelling convention. First we
relate the expansion of Eq.(\ref{kifejt}) in terms of the $20$
komplex numbers $\psi_{\mu\nu\varrho}$ and the $20$ komplex
numbers of the CI expansion in terms of the coefficients of the
"C"s \beq 1=\psi_{123},\qquad
{C_a}^i=\frac{1}{2}\varepsilon^{ijk}\psi_{jka},\qquad
{C_{ab}}^{ij}=\varepsilon^{ijk}\psi_{abk},\qquad
{C_{\overline{123}}}^{123}=\psi_{\overline{123}}. \label{sixC}
\eeq \noindent For notational simplicity we rename the $20$
numbers of this $1+9+9+1$ split in terms of two complex numbers
$\alpha,\beta$  and two $3\times 3$ matrices $A,B$ as follows \beq
\alpha=1,\qquad
{A^a}_i=\frac{1}{4}\varepsilon^{abc}\varepsilon_{ijk}{C_{bc}}^{jk},\qquad
{B^i}_a= {C_a}^i,\qquad \beta={C_{\overline{123}}}^{123}.
\label{sixAB} \eeq \noindent Similarly for the CCparameters we
introduce yet another $1+9+9+1$ split featuring the complex
numbers $\eta,\xi$ and the $3\times 3$ matrices $X,Y$ \beq
\eta=1,\qquad
{X^a}_i=\frac{1}{4}\varepsilon^{abc}\varepsilon_{ijk}{T_{bc}}^{jk},\qquad
{Y^i}_a= {T_a}^i,\qquad \xi={T_{\overline{123}}}^{123}.
\label{sixXY} \eeq \noindent In order to complete the dictionaries
between the different expansions of
Eq.(\ref{kifejt}),(\ref{ccexp}), and (\ref{fci}) we need to relate
the sets $(\alpha,A,B,\beta)$ and $(\eta,X,Y,\xi)$.

 In order to do this let us also compare expansions (\ref{ccexp}) and (\ref{fci}).
Then we get \beq \hat{C}_1=\hat{T}_1,\qquad
\hat{C}_2=\hat{T}_2+\frac{1}{2}\hat{T}_1^2,\qquad
\hat{C}_3=\hat{T}_3+\hat{T}_1\hat{T}_2+\frac{1}{6}\hat{T}_1^3
\label{kapcs1} \eeq \noindent \beq \hat{T}_1=\hat{C}_1,\qquad
\hat{T}_2=\hat{C}_2-\frac{1}{2}\hat{C}_1^2,\qquad
\hat{T}_3=\hat{C}_3-\hat{C}_1\hat{C}_2+\frac{1}{3}\hat{C}_1^3.
\label{kapcs2} \eeq \noindent One can check that \beq
\hat{T}_1\hat{T}_2\hat{p}^1\hat{p}^2\hat{p}^3\vert 0\rangle= {\rm
Tr}(XY)\hat{p}^{\overline{1}}
\hat{p}^{\overline{2}}\hat{p}^{\overline{3}}\vert 0\rangle
\label{1for} \eeq \noindent \beq
\frac{1}{6}\hat{T}_1^3\hat{p}^1\hat{p}^2\hat{p}^3\vert 0\rangle=
({\rm Det}Y) \hat{p}^{\overline{1}}
\hat{p}^{\overline{2}}\hat{p}^{\overline{3}}\vert 0\rangle
\label{2for} \eeq \noindent \beq \frac{1}{2}\hat{T}_1^2
\hat{p}^1\hat{p}^2\hat{p}^3\vert 0\rangle=
{(Y^{\sharp})^a}_i\varepsilon_{abc}\hat{p}^i\hat{p}^b\hat{p}^c\vert 0\rangle
\eeq\noindent \beq
\hat{T}_3\hat{p}^1\hat{p}^2\hat{p}^3\vert 0\rangle= \xi
\hat{p}^{\overline{1}}
\hat{p}^{\overline{2}}\hat{p}^{\overline{3}}\vert 0\rangle,
\label{4for} \eeq \noindent where ${Y}^{\sharp}$ is the adjoint
matrix of $Y$ i.e. \beq YY^{\sharp}=({\rm Det}Y)I \label{adjoint}
\eeq \noindent
 with $I$ the $3\times 3$ identity matrix. Using
these results by virtue of Eq.(\ref{kapcs1}) we obtain the
following dictionary between the CC and CI pictures \beq
\alpha=\eta =1,\qquad B=Y,\qquad A=Y^{\sharp}+X,\qquad \beta ={\rm
Det}Y+(X,Y)+\xi \label{hiperfontos} \eeq \noindent where \beq
(X,Y)\equiv {\rm Tr}(XY). \eeq \noindent

\subsection{The quartic invariant}

It is well-known that we have a quartic combination
$\mathcal{D}(\psi)$ of the $20$ amplitudes of $\vert\psi\rangle$
which is invariant under determinant one SLOCC
transformations\cite{LevVran}, i.e. under the action of the group
$SL(6,\mathbb{C})$ of the form \beq \psi_{\mu\nu\rho}\mapsto
{S_{\mu}}^{\mu^{\prime}}
{S_{\nu}}^{\nu^{\prime}}{S_{\rho}}^{\rho^{\prime}}\psi_{\mu^{\prime}\nu^{\prime}
\rho^{\prime}},\qquad S\in SL(6,\mathbb{C}). \label{SLOCC} \eeq
\noindent This invariant gives rise to a natural measure of
fermionic entanglement which for embedded three-qubit states boils
down\cite{LevVran} to the three-tangle describing the phenomenon
of entanglement monogamy\cite{CKW}. In order to introduce this
invariant first we define the covariant\cite{Hitchin,SarLev2} \beq
{K^{\mu}}_{\nu}=\frac{1}{12}\varepsilon^{\mu\rho_1\rho_2\rho_3\rho_4\rho_5}
\psi_{\nu\rho_1\rho_2}\psi_{\rho_3\rho_4\rho_5} \label{Kcovar} \eeq
\noindent which is transforming under SLOCC as \beq
{K^{\mu}}_{\nu}\mapsto ({\rm Det}S) {S^{\mu}}_{\mu^{\prime}}
{{S^{\prime}_\nu}}^{ \nu^{\prime}}{K^{\mu^{\prime}}}_{\nu^{\prime}}
\label{kkovtrans} \eeq where $S^{\prime}=S^{-1T}$. Then \beq
\mathcal{D}(\psi)=\frac{1}{6}{\rm
Tr}(K^2)=\frac{1}{6}{K^{\mu}}_{\nu} {K^{\nu}}_{\mu}.
\label{kvartik6invar} \eeq \noindent Hence it follows that
$\mathcal{D}(\psi)$ is a relative invariant under SLOCC i.e. \beq
\mathcal{D}(\psi)\mapsto({\rm Det S})^2\mathcal{D}(\psi)
\label{invarprop} \eeq\noindent and invariant under the
$SL(6,\mathbb{C})$ SLOCC subgroup.

We will need the expression for $\mathcal{D}(\psi)$ in the
parametrization of Eq.(\ref{sixAB}). It is given by the
formula\cite{LevVran} \beq
\mathcal{D}(\psi)=4[\kappa^2-(A^{\sharp},B^{\sharp})+\alpha{\rm
Det}A+\beta{\rm Det}B] ,\qquad 2\kappa=\alpha\beta-(A,B).
\label{kvartik} \eeq \noindent This expression displays the
parameters $(\alpha,A,B,\beta)$ i.e. the ones of the full CI
expansion of $\vert\psi\rangle$ of (\ref{fci}). Using Eq.
(\ref{hiperfontos}) let us now give its new expression in terms of
the CC expansion parameters $(\eta,Y,X,\xi)$. The result is the
much simpler expression \beq \mathcal{D}(\psi)=\xi^2+4{\rm Det}X.
\label{kvartik2} \eeq \noindent Hence this expression is not
featuring the matrix $Y$ at all!

Looking back at Eq.(\ref{kvartik}) we see that the alternative
choice \beq \alpha^{\prime} =1,\qquad B^{\prime}=0,\qquad
A^{\prime}=X,\qquad \beta^{\prime} =\xi \label{hiperfontos2} \eeq
\noindent obtained by setting $Y=0$ in Eq.(\ref{hiperfontos})
immediately yields Eq.(\ref{kvartik2}). Hence we suspect that the
states $\vert\psi\rangle$ answering the set $(\alpha,A,B,\beta)$
and $\vert\psi^{\prime}\rangle$ answering the one
$(\alpha^{\prime},A^{\prime},B^{\prime},\beta^{\prime})$ are
related by a special $SL(6,\mathbb{C})$ transformation.

This is indeed the case.
The group $SL(6, \mathbb{C})$ is having a subgroup of upper triangular matrices
of the form
\beq
\begin{pmatrix}
I&\Lambda\\0&I\end{pmatrix},\qquad \Lambda\in Matr(3,\mathbb{C}).
\label{trianglematr}\eeq \noindent  In the language of the set
$(\alpha,A,B,\beta)$ these transformations are of the
form\cite{LevVran} \beq \alpha^{\prime}=\alpha,\qquad
B^{\prime}=B+\alpha\Lambda,\qquad A^{\prime}=A+2B\times
\Lambda+\alpha \Lambda^{\sharp} \label{Btrans1} \eeq \noindent
\beq
\beta^{\prime}=\beta+(A,\Lambda)+(B,\Lambda^{\sharp})+\alpha{\rm
Det}\Lambda \label{Btrans2} \eeq \noindent where \beq
2B\times
\Lambda=(B+\Lambda)^{\sharp}-B^{\sharp}-\Lambda^{\sharp}.
\label{formula1} \eeq \noindent Now we can see that with the
choice \beq \Lambda\equiv -Y \label{bpara} \eeq \noindent
 from the set of
Eq.(\ref{hiperfontos}) we can obtain the one of
Eq.(\ref{hiperfontos2}). The upshot of these considerations is
that using a SLOCC transformation corresponding to the matrix of
Eq.(\ref{trianglematr}) with parameter given by (\ref{bpara}) we
can get rid of the $\hat{T}_1$ term in the CC expansion of
Eq.(\ref{ccexp}). Hence in order to reveal the interplay between
the CCM and the SLOCC classes it is sufficient to examine the
structure of the state \beq \vert\psi^{\prime}\rangle
=e^{\hat{T}_2+\hat{T}_3}\vert\psi_0\rangle \label{psiv} \eeq
\noindent with parameters given by (\ref{hiperfontos2}).

It is easy to see that this important conclusion is valid also in
the most general case (i.e. the number of modes can be arbitrary,
say $N$). Indeed, since the operators $\hat{T}_1,\hat{T}_2$ and
$\hat{T}_3$ are commuting ones we can write \beq
\vert\psi\rangle=e^{\hat{T}_1}e^{\hat{T}_2}e^{\hat{T}_3}\vert\psi_0\rangle.
\label{szepalak} \eeq \noindent
Now it is easy to check that the
operator $e^{\hat{T}_1}$ always generates SLOCC transformations of
upper triangular form. To cap all this our observation is valid
for an arbitrary number of (say $n$) fermions so one can write
\beq \vert\Psi\rangle =e^{\hat{T}_1}\vert{\Psi}^{\prime}\rangle,
\qquad\vert{\Psi}^{\prime}\rangle=e^{\hat{T}_2}\cdots
e^{\hat{T}_n}\vert\Psi_0\rangle,\qquad e^{\hat{T}_1}\in
GL(N,\mathbb{C}) \label{vesszosalak} \eeq \noindent 
where
\beq
\vert\Psi_0\rangle\equiv\hat{p}^1\hat{p}^2\cdots\hat{p}^n\vert 0\rangle.
\eeq
\noindent
Hence again
$\vert\Psi\rangle$ and $\vert{\Psi}^{\prime}\rangle$ are in the
same SLOCC class. In summary: in the CC picture SLOCC entanglement
is characterized merely by the cluster operators
$\hat{T}_2,\dots,\hat{T}_n$. Then the CC expansion featuring only
these operators represent deviations from the separable class
represented by a single Slater determinant i.e. a HF state of the
form \beq \vert\Psi_B\rangle\equiv
e^{\hat{T}_1}\vert\Psi_0\rangle. \label{Bruckner} \eeq \noindent
The state $\vert\Psi_B\rangle$ can also be written as a single
Slater determinant in a suitable basis.
In the case of our example of three fermions with six modes we have
\beq
\vert\psi_B\rangle=e^{\hat{T}_1}\vert\psi_0\rangle=(\hat{p}^1+{Y^1}_a\hat{p}^a)
(\hat{p}^2+{Y^2}_b\hat{p}^b)
(\hat{p}^3+{Y^3}_c\hat{p}^c)\vert 0\rangle.
\label{Brucknerbase}
\eeq
\noindent

\subsection{SLOCC classes and CCM}

As described elsewhere\cite{LevVran} given a state
$\vert\psi\rangle$ one can introduce a dual state
$\vert\tilde{\psi}\rangle$ which is a cubic expression of the
original amplitudes of $\vert\psi\rangle$ as follows \beq
\vert\tilde{\psi}\rangle\longleftrightarrow
(\tilde{\alpha},\tilde{A},\tilde{B}, \tilde{\beta})
\label{mapping} \eeq \noindent
\beq
\tilde{\alpha}=2\alpha\kappa+2{\rm Det}B,\qquad
\tilde{A}=2(\beta B^{\sharp}-2B\times A^{\sharp})-2\kappa A.
\label{dua1}
\eeq
\noindent
\beq
\tilde{\beta}=-2\beta\kappa-2{\rm Det}A,\qquad
\tilde{B}=-2(\alpha A^{\sharp}-2A\times B^{\sharp})+2\kappa B,
\label{dua2}
\eeq
\noindent
where $\kappa$ and the cross product are defined in Eqs.(\ref{kvartik})
and (\ref{formula1}).
One can also introduce the following set of quadratic polynomials
\beq Q_1=\alpha\beta I-AB,\qquad Q_2\equiv A^{\sharp}-\beta
B,\qquad Q_3\equiv B^{\sharp}-\alpha A. \label{plucker} \eeq
\noindent In fact these $3\times 3$ matrices are just coming from
the blocks of the covariant of Eq.(\ref{Kcovar}) hence $K_{\psi}=0$ iff
$Q_1=Q_2=Q_3=0$.
Then it is known\cite{LevVran,SarLev2} that the SLOCC classes can
be characterized as follows.

Let us denote by $\mathcal{M}_0$ our space of three fermions with six modes.
Clearly $\mathcal{M}_0\simeq\wedge^3\mathbb{C}^6$.
We make the further definitions
\beq
\mathcal{M}_1\equiv\{\psi\in\mathcal{M}_0 \vert \mathcal{D}(\psi)=0\}
\label{m1}
\eeq
\noindent
\beq
\mathcal{M}_2\equiv\{\psi\in\mathcal{M}_0 \vert \mathcal{D}(\psi)=0, \tilde{\psi}=0\}
\label{m2}
\eeq
\noindent
\beq
\mathcal{M}_3\equiv\{\psi\in\mathcal{M}_0\vert  K_{\psi}=0\}
\label{m3}
\eeq
\noindent
Notice that in the case of $\mathcal{M}_3$ the vanishing of the covariant $K_{\psi}$ automatically implies
$\mathcal{D}(\psi)=0, \tilde{\psi}=0$.
Now we have\cite{LevVran} four nontrivial $GL(6,\mathbb{C})$ orbits which we call SLOCC entanglement classes
\beq
\mathcal{O}_{GHZ}\equiv \mathcal{M}_0-\mathcal{M}_1=[(\hat{p}^{123}+\hat{p}^{1\overline{2}\overline{3}}
+\hat{p}^{\overline{1}2\overline{3}}+\hat{p}^{\overline{1}\overline{2}3}\vert 0\rangle)]
\label{GHZcl}
\eeq
\noindent
\beq
\mathcal{O}_{W}\equiv \mathcal{M}_1-\mathcal{M}_2
=
[(\hat{p}^{123}+\hat{p}^{1\overline{2}\overline{3}}
+\hat{p}^{\overline{1}2\overline{3}}\vert 0\rangle)]
\label{Wcl}
\eeq
\noindent
\beq
\mathcal{O}_{BISEP}\equiv \mathcal{M}_2-\mathcal{M}_3
=[(\hat{p}^{123}+\hat{p}^{1\overline{2}\overline{3}}
\vert 0\rangle)]
\label{BISEPcl}
\eeq
\noindent
\beq
\mathcal{O}_{SEP}\equiv \mathcal{M}_3=[\hat{p}^{123}\vert 0\rangle]
\label{SEPcl}
\eeq
\noindent
where the notation $[\vert\psi\rangle]$ refers to the SLOCC orbit (equivalence class) of the unnormalized state $\vert\psi\rangle$

The SEP (separable) class is the class consisting of states that can be written as a single Slater determinant.
The BISEP (biseparable) class corresponds to the set of tripartite states that can be transformed via SLOCC to states
consisting of the sum of two Slater determinants containing a common factor of single particle states.
The W and GHZ classes form the genuine entanglement classes.
The names separable, biseparable, W and GHZ are originating from the fact that when
considering embedded three-qubit systems living inside our three fermionic ones\cite{LevVran,Chen0} these classes correspond to the separable, biseparable, W and GHZ classes known from studies of three-qubit entanglement\cite{Dur}.

Since $\mathcal{ D}$ is a $SL(6,\mathbb{C})$ invariant and
$\vert\tilde{\psi}\rangle$ and $K$ are covariants one can
calculate these quantities merely for the transformed
representative $\vert\psi^{\prime}\rangle$ also belonging to the
orbit of $\vert\psi\rangle$. Using Eq.(\ref{hiperfontos2}) one
obtains
\beq
\tilde{\alpha}^{\prime}=\xi,\qquad {\tilde{A}}^{\prime}=-\xi X,\qquad {\tilde{B}}^{\prime}=-2X^{\sharp},\qquad
\tilde{\beta}^{\prime}=-\xi^2-2{\rm Det}X
\label{expltrilifr2} \eeq \noindent \beq Q^{\prime}_1=\xi I, \qquad
Q^{\prime}_2=X^{\sharp}, \qquad Q^{\prime}_3=-X. \label{constrpl} \eeq \noindent
From these results we see that:

{\bf I.}  The matrix $Y$ plays no role in the SLOCC
classes.

{\bf II.} The SEP (single Slater determinant) states are {\bf
precisely} the states characterized by the constraints $\xi=0$ and
$X=0$. For these states $\mathcal{D}$, $\vert\tilde{\Psi}\rangle$
and $K_{\psi}$ are vanishing automatically.

{\bf III.}  The BISEP states are characterized by the constraints
$\xi=0$, $X^{\sharp}=0$ , {\rm Det}X=0 but $X\neq 0$ . These constraints
automatically yield $\mathcal{D}=0$ and
$\vert\tilde{\Psi}\rangle=0$, however $K_{\psi}\neq 0$.

{\bf IV.}  The W states are characterized by the constraint
$\xi^2+4{\rm Det}X =0$,  and the nonvanishing of 
{\it at least one} of the quantities of Eq.(\ref{expltrilifr2}).
So for instance we are in the W-class if we have $\xi=0, X\neq 0, X^{\sharp}\neq 0, {\rm Det}X=0$.

{\bf V.}  The GHZ states are characterized by the constraint
$\xi^2+4{\rm Det}X\neq 0$. These states are {\it dense} within the
full set of states.

Recall that $\xi$ and $X$ characterize
the contributions coming from triples and doubles of
cluster operators.
From the above results we also see that in order to find out which entanglement class a state belongs the derived quantities of $X^{\sharp}$ and ${\rm Det}X$ should also to be taken into consideration.
Summarizing: interestingly the notion of a SLOCC transformation shows up quite naturally
as a one relating the CC and CI expansions.
Going from the CI expansion to the CC one seems to play the role of obtaining
something like a simplified form for the three fermion state where the structure of the entanglement classes is transparent.
In order to see how our observations generalize in the next section we turn to a more complicated system.

\section{Three fermions with seven single particle states}

\subsection{Invariants and covariants}

A three fermion state with seven modes can be written in the
following form

\beq \vert\Psi\rangle=\frac{1}{3!}\Psi_{IJK}\hat{p}^{IJK}\vert
0\rangle \label{7psi} \eeq \noindent where $I,J,K=1,2,\dots 7$ and
$\Psi_{IJK}$ is a totally antisymmetric complex tensor having $35$
independent components. The SLOCC group is $GL(7,\mathbb{C})$
and the group action is the usual one \beq \Psi_{IJK}\mapsto
{S_I}^{I^{\prime}}
{S_J}^{J^{\prime}}{S_K}^{K^{\prime}}\Psi_{I^{\prime}J^{\prime}K^{\prime}},\qquad
S\in GL(7,\mathbb{C}). \label{traf7} \eeq\noindent

 The basic
covariants are\cite{SarLev2} \beq
{(M^I)^J}_K=\frac{1}{12}\varepsilon^{IJA_1A_2A_3A_4A_5}\Psi_{KA_1A_2}\Psi_{A_3A_4A_5}
\label{Mkovar} \eeq \noindent\beq
N_{IJ}=\frac{1}{24}\varepsilon^{A_1A_2A_3A_4A_5A_6A_7}\Psi_{IA_1A_2}\Psi_{JA_3A_4}\Psi_{A_5A_6A_7}
\label{Nkovar} \eeq \noindent\beq
L^{IJ}={(M^I)^{A_1}}_{A_2}{(M^J)^{A_2}}_{A_1}. \label{Lkovar} \eeq
\noindent Under SLOCC the transformation properties of these
covariants are as follows \beq {(M^I)^J}_K\mapsto ({\rm Det}
S){S^I}_{I^{\prime}}
{S^J}_{J^{\prime}}{S^{\prime}_K}^{K^{\prime}}{(M^{I^{\prime}})^{J^{\prime}}}_{K^{\prime}}
\label{Mkovtr} \eeq \noindent \beq N_{IJ}\mapsto ({\rm Det}S)
{S^{\prime}_I}^{I^{\prime}} {S^{\prime}_J}^{J^{\prime}}
N_{I^{\prime}J^{\prime}} \label{Nkovtr} \eeq\noindent \beq
L^{IJ}\mapsto ({\rm Det}S)^2
{S^I}_{I^{\prime}}{S^J}_{J^{\prime}}L^{I^{\prime}J^{\prime}}
\label{Lkovtr} \eeq \noindent where $S^{\prime}=(S^{-1})^T$. Notice
that the $7\times 7$ matrices $N_{IJ}$ and $L^{IJ}$ are symmetric.

From these covariants one can form a unique algebraically
independent relative invariant \beq
\mathcal{J}(\Psi)=\frac{1}{2^4\cdot 3^2\cdot 7}{\rm Tr}(NL)
=\frac{1}{2^4\cdot 3^2\cdot 7}N_{IJ}L^{IJ}.
\label{jeinv}\eeq\noindent $\mathcal{J}$ is a relative invariant
meaning that under SLOCC it picks up a determinant factor \beq
\mathcal{J}(\Psi)\mapsto ({\rm Det} S)^3\mathcal{J}(\Psi)
\label{transje} \eeq hence it is invariant under the SLOCC
subgroup $SL(7,\mathbb{C})$. Defining \beq
\mathcal{B}_{IJ}=-\frac{1}{6}N_{IJ}\label{calbe} \eeq\noindent we
have the alternative formula \beq (\mathcal{J}(\Psi))^3={\rm
Det}\mathcal{B}. \label{kobinv} \eeq \noindent

\subsection{Relating the CC and CI expansions for seven modes}

Similarly to the six mode case for a coupled cluster description
of this three fermion system we split the modes to ones that are
occupied labelled by $i,j,k=1,2,3$ and the ones that are not
occupied by
$a,b,c=\overline{1},\overline{2},\overline{3},\overline{4}$. Now
the CC and CI expansions will be just the same form as the ones
displayed in Eqs.(\ref{HFstate})-(\ref{tkc}) with the exception of
$\hat{T}_3$ and $\hat{C}_3$ having the new form \beq
\hat{T}_3=\frac{1}{3!}{T_{abc}}^{123}\hat{p}^a\hat{n}_1\hat{p}^b\hat{n}_2\hat{p}^c\hat{n}_3,
\qquad
\hat{C}_3=\frac{1}{3!}{C_{abc}}^{123}\hat{p}^a\hat{n}_1\hat{p}^b\hat{n}_2\hat{p}^c\hat{n}_3.
\label{newTC} \eeq \noindent

It is rewarding to tackle this case as a deviation from the six
mode one. We write
\beq
\vert\Psi\rangle=\vert\psi\rangle+\vert\omega\rangle,\qquad \vert\omega\rangle =\frac{1}{2}\omega_{\mu\nu}\hat{p}^{\mu\nu\overline{4}}
\vert 0\rangle
\label{kahlerdecomp}
\eeq
\noindent
and $\vert\psi\rangle$ is given by Eq.(\ref{kifejt}). The $6\times 6$ antisymmetric matrix $\omega$ has 
$15$ independent components. We write
\beq
\omega\equiv\begin{pmatrix}E&D\\-D^T&F\end{pmatrix}
\label{omegakiir}
\eeq
\noindent
where $E,F$ are two $3\times 3$ antisymmetric matrices and $D$ is an arbitrary $3\times 3$ matrix. 
Hence in the CI and CC pictures we group the $35$
amplitudes to three $3\times 3$ matrices, and two 
$3\times 3$ antisymmetric ones so in the CI
picture we will have five matrices
$A,B,D,E,F$ with
the latter two being antisymmetric, and two scalars $\alpha$ and
$\beta$. Similarly in the CC picture the five matrices will be
denoted as $X,Y,Z,U,V$ with the latter two antisymmetric, and the
corresponding scalars are $\eta$ and $\xi$. The scalars and the
first two matrices in both pictures have already been defined by
Eqs.(\ref{sixAB}) and (\ref{sixXY}).
The remaining matrices are defined as
\beq
{C_{a\overline{4}}}^{ij}=\varepsilon^{ijk}\Psi_{ka\overline{4}}
=
\varepsilon^{ijk}D_{ka}
,\qquad
{C_{\overline{4}}}^i=\frac{1}{2}\varepsilon^{ijk}\Psi_{jk\overline{4}}
=
\frac{1}{2}\varepsilon^{ijk}E_{jk}
\label{ujCI1}
\eeq
\noindent
\beq
 {C_{a
b\overline{4}}}^{123}=\Psi_{ab\overline{4}}=F_{ab}
.
\label{ujCI2}
\eeq
\noindent
\beq
{T_{a\overline{4}}}^{ij}=
\varepsilon^{ijk}Z_{ka}
,\qquad
{T_{\overline{4}}}^i=
\frac{1}{2}\varepsilon^{ijk}V_{jk},\qquad
 {T_{a
 b\overline{4}}}^{123}=U_{ab}
 \label{ujCc}
 \eeq
 \noindent
where in these expressions we have $a,b=\overline{1},\overline{2},\overline{3}$.

Now similarly to the six mode case we expect that the form of the unique relative invariant $\mathcal{J}(\Psi)$ of Eq.(\ref{jeinv})  will be  of much simpler form in the CC than in the CI picture.
Indeed, in the next subsection it turns out that unlike in the CI picture where this invariant is featuring all of the amplitudes
$(\alpha, \beta, A,B,D,E,F)$ in the CC one only the quantities
$(\eta=1,\xi, X,Z,U)$ show up.

In order to prove this again we have to relate the CI and CC amplitudes using Eq.(\ref{kapcs1}).
Then a calculation shows that
\beq
\alpha=\eta=1,\qquad \beta=\xi+{\rm Tr}(XY)+{\rm Det}Y,\qquad B=Y,\qquad A=X+Y^{\sharp}
\label{elso7mode}
\eeq
\noindent
\beq
D=Z+VY,\qquad E=V, \qquad F=U+(Z^TY-Y^TZ)+[(X+Y^{\sharp})v].
\label{masodik7mode}
\eeq
\noindent
Here
\beq
v^i=\frac{1}{2}\varepsilon^{ijk}V_{jk}, \qquad V_{ij}\equiv [v]_{ij}=\varepsilon_{ijk}v^k
\label{matrixvector}
\eeq
\noindent
i.e. the latter notation refers to building up a $3\times 3$ antisymmetric matrix from a $3$ component vector.

\subsection{The seventh order invariant}

The data $(\alpha, \beta, A,B,D,E,F)$ described in the previous section is directly related to the complex amlitudes $\Psi_{IJK}$. They can be used for calculating the covariants and the invariant of Eq.(\ref{jeinv}).
Again the data connected to the operator $\hat{T}_1$ is not needed since it can be linked to a SLOCC transformation of upper triangular form. It means that it is enough to do the calculation with $V=Y=0$.
The result for the calculation of the components of the covariants $N_{IJ}$ and $M^{IJ}$ is given in the Appendix. Putting everything together the final result is the surprisingly simple expression
\beq
\mathcal{J}(\Psi)=-{\rm Det}(G)-\frac{1}{4\xi}{\rm Det}(UX+\xi Z^T),\qquad
G\equiv \frac{1}{2}(ZX+X^TZ^T),
\label{klassz}
\eeq
\noindent
where the matrix $G$ is just the symmetric part of the $3\times 3$ matrix $ZX$.
Notice that the expression is regular for $\xi=0$ since for arbitrary $3\times 3$ matrices we have
\beq
{\rm Det}(A+B)={\rm Det}A+{\rm Tr}(A^{\sharp}B)+{\rm Tr}(AB^{\sharp})+{\rm Det}B,
\label{identity}
\eeq
\noindent
and ${\rm Det}UX=0$ due to $U^T=-U$.

Let us find inside this formula our invariant
of Eq.(\ref{kvartik2})
familiar from the six mode case!
First notice that due to Eq.(\ref{matrixvector}) and the antisymmetry of $U$ we have $uu^T=U^{\sharp}$.
Then introducing yet another $3$ component vector $w$ and its corresponding antisymmetric matrix
$W\equiv [w]$ 
as
\beq
W\equiv \frac{1}{2}(ZX-X^TZ^T),\qquad
w^i=\frac{1}{4}\varepsilon^{ijk}(ZX-X^TZ^T)_{jk}
\label{antikapart}
\eeq
\noindent
we obtain
\beq
\mathcal{J}(\Psi)=w^TGw-\frac{1}{4}u^THu-\frac{1}{4}\left({\rm Det}Z(\xi^2+4{\rm Det} X)\right)-\frac{1}{4}\xi{\rm Tr}(UXZ^{T\sharp}),
\label{klasszezis}
\eeq
\noindent
\beq
H=\frac{1}{2}(Z^TX^{\sharp}+X^{\sharp T}Z).
\eeq
\noindent
Here we also used that
\beq
{\rm Det}(ZX)={\rm Det}(W+G)={\rm Det}G+Tr(W^{\sharp}G)={\rm Det}G+w^TGw.
\eeq
\noindent
From Eq.(\ref{klassz}) it is also clear that 
if we have $G=ZX$ (i.e. when the matrix $ZX$ is symmetric) and $U=0$
then
$\mathcal{J}(\Psi)=-{\rm Det}Z(\xi^2+4{\rm Det}X)/4$.

An equivalent way of formulating this can be given as follows.
Define $\hat{\omega}=\frac{1}{2}\omega_{\mu\nu}\hat{p}^{\mu\nu\overline{4}}$ and apply the restricion $Y=V=0$.
Then according to Eq.(\ref{masodik7mode}) then $D=Z$, $E=0$ and $F=U$ 
hence one can verify that
\beq
\hat{\omega}\vert\psi\rangle=0 \qquad {\rm iff}\qquad G=ZX,\quad U=0.
\label{compatib}
\eeq
\noindent
Recall that ${\rm Det}\omega =({\rm Pf}\omega)^2$
where
\beq
{\rm Pf}(\omega)=\frac{1}{2^33!}\varepsilon^{\mu_1\mu_2\mu_3\mu_4\mu_5\mu_6}\omega_{\mu_1\mu_2}\omega_{\mu_3\mu_4}\omega_{\mu_5\mu_6}
\label{Pfaffian}
\eeq
\noindent
then in the case when $V=U=Y=0$ we have ${\rm Det Z}=-{\rm Pf}\omega$.  
Using this one obtains the result
\beq
\mathcal{J}(\Psi)=\frac{1}{4}{\rm Pf}(\omega)\mathcal{D}(\psi).
\label{faktoriz}
\eeq
\noindent
It can be proved that this result remains valid also in the case when the restrictions $V=Y=0$ are removed\cite{SarLev2}.
Hence $\hat{\omega}\vert\psi\rangle=0$ of Eq.(\ref{compatib}) can be regarded as a necessary condition for the separability of the seventh order invariant to a cubic and a quartic one.

\subsection{Entanglement classes}

The SLOCC classification of three-fermion states with seven modes is known in the mathematical literature
as the classification problem of three-forms in a seven dimesional vector space\cite{Reichel,Schouten, Gurevich}.
These results has recently been introduced to entanglement theory\cite{SarLev2}.
Here we would like to recall the structure of the SLOCC classes.

Let us consider the state
\beq
\vert\Psi_-\rangle=(\hat{p}^{123}-\hat{p}^{1\overline{2}\overline{3}}-
\hat{p}^{2\overline{3}\overline{1}}-\hat{p}^{3\overline{1}\overline{2}}+\hat{p}^{1\overline{1}\overline{4}}
+\hat{p}^{2\overline{2}\overline{4}}+\hat{p}^{3\overline{3}\overline{4}})\vert 0\rangle.
\label{kanghz71}
\eeq
\noindent
In this special case due to relations
(\ref{elso7mode})-(\ref{masodik7mode})
we have
\beq
(\alpha,A,B,\beta,D,E,F)=(\eta,X,Y,\xi,Z,V,U)=(1,-I,0,0,I,0,0).
\label{set1}
\eeq
\noindent
(Notice that according to our findings at the end of Section 2.2. one can always achieve that $V=Y=0$
and in this case by virtue of Eqs.(\ref{elso7mode}) and (\ref{masodik7mode}) the CC and CI labels are just the same.)
Using now Eq.(\ref{klassz}) with $Z=-X=I$, and $\xi=U=0$ we obtain
\beq
\mathcal{J}(\Psi_-)=1.
\eeq
\noindent
It can be shown\cite{SarLev2} that this three-fermion state with seven modes belongs to a dense SLOCC orbit characterized by the property $\mathcal{J}(\Psi_-)\neq 0$. Notice that from the mathematical point of view this orbit is similar to the dense SLOCC orbit of the famous GHZ state of the six mode case for which we had $\mathcal{D}\neq 0$. 
Indeed apart from some signs the first four terms of Eq.(\ref{kanghz71}) correspond to a structure already
known from Eq.(\ref{GHZcl}).

In order to make the connection with the three-qubit GHZ-state more explicit we
introduce the complex linear combinations
\beq
\hat{E}^{1,2,3}=\hat{p}^{1,2,3}+i\hat{p}^{\overline{1},\overline{2},\overline{3}},\qquad
\hat{E}^{\overline{1},\overline{2},\overline{3}}=\hat{p}^{1,2,3}-i\hat{p}^{\overline{1},\overline{2},\overline{3}}
,\qquad \hat{E}^{\overline{4}}=i\hat{p}^{\overline{4}}.
\label{ujkombi}
\eeq
\noindent
Then one can write
\beq
\left(\hat{p}^{123}-\hat{p}^{1\overline{2}\overline{3}}-\hat{p}^{\overline{1}2\overline{3}}-
\hat{p}^{\overline{1}\overline{2}3}\right)\vert 0\rangle=\frac{1}{2}\left(\hat{E}^{123}+\hat{E}^{\overline{123}}\right)\vert 0\rangle.
\label{ghzsalak}
\eeq
\noindent
Then the state on the right hand side of Eq.(\ref{ghzsalak}) corresponds to the structure $\vert 000\rangle +\vert 111\rangle$ of the unnormalized three-qubit GHZ state.
It can be proved that the correpondence is not superficial since the three-qubit SLOCC group can be embedded in
a consistent manner to the fermionic SLOCC one\cite{LevVran,Chen0}.
Now our state of Eq.(\ref{kanghz71}) can be given the following form
\beq
\vert\Psi_-\rangle=\frac{1}{2}\left(\hat{E}^{123}+\hat{E}^{\overline{123}}+(\hat{E}^{1\overline{1}}+
\hat{E}^{2\overline{2}}+\hat{E}^{3\overline{3}})\hat{E}^{\overline{4}}\right)\vert 0\rangle.
\label{ezazujalakja}
\eeq
\noindent
The reason for the notation $\vert\Psi_-\rangle$ corresponds to the fact that according to the left hand side of Eq.(\ref{ghzsalak}) and Eq.(\ref{kvartik2}) the $\mathcal{D}$ invariant of the GHZ-part is negative (see also Eq.(\ref{faktoriz}) and the preceeding discussion).

Now introducing the notation
\beq
\vert\Psi\rangle\equiv\hat{\Psi}\vert 0\rangle
\label{vakumra}
\eeq
\noindent
the SLOCC entanglement classes are given in Table 1.
We note that for a full characterization of these classes other covariants are also needed\cite{SarLev2}.
However, for our purposes it is sufficient to consider merely the quantities $N_{IJ}$ and $\mathcal{J}$.
Notice also that according to Eq.(\ref{omegakiir}) the $6\times 6$ antisymmetric matrix defines an alternating
(symplectic) form on the six dimensional space of single particle states. Then the operator
$\hat{E}^{1\overline{1}}+
\hat{E}^{2\overline{2}}+\hat{E}^{3\overline{3}}$ defines a canonical form for a nondegenerate symplectic form.
This explains the abbreviation SYMPL showing up in the second column of our table.
The other abbreviations conform with our ones familiar from Eqs.(\ref{GHZcl})-(\ref{SEPcl}).

\begin{table}[h!]
\centering
\begin{tabular}{|c|c|c|c|c|}
\hline Name & Type & Canonical form of $\hat{\Psi}$ & Rank
$N_{IJ}(\Psi)$ & $\mathcal{J}(\Psi)$ 
\\ \hline \hline
I & NULL & 0 & 0 & 0\\
II & SEP & $\hat{E}^{123}$ & 0 & 0\\
III & BISEP & $\hat{E}^1(\hat{E}^{23}+\hat{E}^{\bar 2\bar 3})$ & 0 & 0\\
IV & W & $\hat{E}^{12\bar 3}+\hat{E}^{1\bar 2 3} + \hat{E}^{\bar 1 2 3}$ & 0 & 0\\
V & GHZ & $\hat{E}^{123} +\hat{E}^{\bar 1\bar 2\bar 3}$ & 0 & 0\\
VI & SYMPL/NULL & $(\hat{E}^{1\bar 1}+\hat{E}^{2\bar 2}+\hat{E}^{3\bar 3})\hat{E}^{\overline{4}}$ & 1 & 0\\
VII & SYMPL/SEP & $(\hat{E}^{1\bar 1}+\hat{E}^{2\bar 2}+\hat{E}^{3\bar 3})\hat{E}^{\overline{4}} +
\hat{E}^{123}$ & 1 & 0\\
VIII & SYMPL/BISEP & $(\hat{E}^{1\bar 1}+\hat{E}^{2\bar 2}+\hat{E}^{3\bar
3})\hat{E}^{\overline{4}}+\hat{E}^1(\hat{E}^{23}+\hat{E}^{\bar 2\bar 3})$ & 2 & 0\\
IX & SYMPL/W & $(\hat{E}^{1\bar 1}+\hat{E}^{2\bar 2}+\hat{E}^{3\bar 3})\hat{E}^{\overline{4}}+\hat{E}^{12\bar 3}+\hat{E}^{1\bar 2 3} + \hat{E}^{\bar 1 2 3}$ & 4 & 0\\
X & SYMPL/GHZ & $(\hat{E}^{1\bar 1}+\hat{E}^{2\bar 2}+\hat{E}^{3\bar
3})E^{\overline{4}}+\hat{E}^{123} +\hat{E}^{\bar 1\bar 2\bar 3}$ & 7 & $\neq 0$\\
\hline
\end{tabular}
\caption{Entanglement classes of three fermions with seven single particle states.
Representative states of the SLOCC classes are obtained by acting with $\hat{\Psi}$ on the vacuum as in Eq.(\ref{vakumra}). The covariant $N_{IJ}$ and invariant $\mathcal{J}$ are defined by Eqs.(\ref{Nkovar}) and (\ref{jeinv}).}
\label{tab:2}
\end{table}

\subsection{Perturbing the canonical GHZ-like state}

In this section we would like to obtain an insight into the interplay between the structure of cluster operators $\hat{T}_2$ and $\hat{T}_3$ and the entanglement classes of the previous section.
In order to achieve this as a first step notice that in our
description of the SLOCC classes for three-fermions with seven modes  we managed to produce the "GHZ-like" state 
$\vert\Psi_-\rangle$ belonging to the dense orbit merely via an application of cluster operators containing only doubles. This means that when building up this state only the expansion coefficients of $\hat{T}_2$ were used (see Eqs.(\ref{sixXY}), (\ref{ujCc}) and (\ref{set1})).

Now we would like to perturb this state by adding to it a new term which is also containing 
contribution from triples i.e. from the expansion coefficients of the operator $\hat{T}_3$.
To this end  we choose
\beq
\vert\Phi_-\rangle=\vert\Psi_-\rangle +\vert\chi\rangle,\qquad
\vert\chi\rangle=\left(\xi\hat{p}^{\overline{123}}+u^{\overline{1}}\hat{p}^{\overline{234}}
+u^{\overline{2}}\hat{p}^{\overline{314}}+u^{\overline{3}}\hat{p}^{\overline{124}}
\right)\vert 0\rangle.
\label{perturbo}
\eeq
\noindent
Using Eq.(\ref{jeinv}) we obtain
\beq
\mathcal{J}(\Phi_-)=1-\frac{1}{4}\left(\xi^2+(u^{\overline{1}})^2+
(u^{\overline{2}})^2+(u^{\overline{3}})^2\right).
\label{perturbo2}
\eeq
\noindent
Hence we remain in the dense orbit unless the condition
\beq
\xi^2+(u^{\overline{1}})^2+
(u^{\overline{2}})^2+(u^{\overline{3}})^2=4.
\label{conditiondeg}
\eeq
\noindent
holds.
Notice that the equation
\beq
\xi^2+(u^{\overline{1}})^2+
(u^{\overline{2}})^2+(u^{\overline{3}})^2=0.
\label{conifold}
\eeq
\noindent
defines a six dimensional conifold\cite{Candelas}.
Just as a two-dimensional cone is embedded in real three-dimensional space as $x^2+y^2-z^2=0$, a real six dimensional
conifold is embedded in $\mathbb{C}^4$ via Eq.(\ref{conifold}).
The conifold is a smooth space apart from a singularity at $\xi=u^{\overline{1}}=
u^{\overline{2}}=u^{\overline{3}}=0$.
It is known\cite{Candelas} that one way to repair this singularity is given by a process called deformation under which Eq.(\ref{conifold}) is modified to 
\beq
\xi^2+(u^{\overline{1}})^2+
(u^{\overline{2}})^2+(u^{\overline{3}})^2=Q^2
\label{defconi}
\eeq
\noindent
where $Q$ is a real deformation parameter.
Now we see that Eq.(\ref{conditiondeg}) giving the condition needed for leaving the dense SLOCC orbit is
the one of the perturbing parameters coming from triples defining a deformed conifold with the special deformation parameter $Q=2$.

For real parameters $(\xi,u^a)\in \mathbb{R}^4$ this means that if the $4$ parameters corresponding to the cluster operators describing triples are belonging to a three dimensional sphere of radius $2$
then the entanglement type is changed. 
The parametrized family of three fermion states in this case leaves the dense orbit.
From Table. 1. we see that apart from the entanglement class corresponding to the dense SLOCC orbit, we have $9$ more classes\footnote{
Taken together with the trivial class represented by the zero state.}.
As a next step we show how this class can be identified.

In order to do this we calculate the covariant $N_{IJ}$ a symmetric $7\times 7$ matrix and look at how its rank is behaving as we reach the singular locus of $\mathcal{J}(\Phi_-)$. 
Rather than using the matrix $N_{IJ}$ for convenience we use the one $\mathcal{B}_{IJ}=-\frac{1}{6}N_{IJ}$
of Eq.(\ref{calbe}). For the state $\vert\Psi_-\rangle$ this new matrix becomes just the identity matrix, i.e. we have $\mathcal{B}_{IJ}=\delta_{IJ}$.
Using the expressions given in the Appendix for the perturbed state we get
\beq
\mathcal{B}_{IJ}(\Phi_-)=\begin{pmatrix}I&-\frac{1}{2}(\xi I+U)&-\frac{1}{2}u\\-\frac{1}{2}(\xi I-U)&I&0\\
-\frac{1}{2}u^T&0&1\end{pmatrix}.
\label{calbematrixspec2}
\eeq
\noindent
By applying a suitable (generally complex) orthogonal matrix we can diagonalize $\mathcal{B}$
\beq
\mathcal{B}^{\rm diag}=S^T\mathcal{B}S,\qquad S=\frac{1}{\sqrt{2}Q}\begin{pmatrix}
QI&QI&0\\\xi I-U&U-\xi I&\sqrt{2}u\\u^T&-u^T&-\sqrt{2}\xi\end{pmatrix}\in SO(7,\mathbb{C})
\label{ortomatr}
\eeq
\noindent
\beq
\mathcal{B}^{\rm diag}=\begin{pmatrix}\left(1-\frac{1}{2}Q\right)I&0&0\\
0&\left(1+\frac{1}{2}Q\right)I&0\\
0&0&1\end{pmatrix}
\label{diagort}
\eeq
\noindent
where
\beq
Q\equiv\sqrt{\xi^2+(u^{\overline{1}})^2+
(u^{\overline{2}})^2+(u^{\overline{3}})^2}
\label{Qdefi}
\eeq
\noindent
and recall also our conventions of Eq.(\ref{matrixvector}).
From Eq.(\ref{kobinv}) we see that we are in accord with Eq.(\ref{perturbo2}) and also see that our matrix 
$\mathcal{B}$ fails to be of full rank if the (\ref{conditiondeg}) i.e. $Q=2$ condition holds.
From Eq.(\ref{diagort}) we see that in the degenerate case the rank of $\mathcal{B}$ is reduced from seven to four. From Table. 1. we se that this reduction of rank of $\mathcal{B}$ indicates a transition from entanglement class X. to class IX. 
The former class is the class of the state $\vert\Psi_{-}\rangle$ of Eq.(\ref{kanghz71}) and the latter is the SLOCC class of the state
obtained from Eq.(\ref{kahlerdecomp}) by keeping $\vert\omega\rangle$ with $E=F=0$ and $D=I$ and using for $\vert\psi\rangle$ a state from the $W$-class of the Eq.(\ref{Wcl}) form.

As an other example let us consider the state
\beq
\vert\Phi_+\rangle=\vert\Psi_+\rangle +\vert\chi\rangle
\label{plusszos}
\eeq
\noindent
where
\beq
\vert\Psi_+\rangle=(\hat{p}^{123}+\hat{p}^{1\overline{2}\overline{3}}+
\hat{p}^{2\overline{3}\overline{1}}+\hat{p}^{3\overline{1}\overline{2}}+\hat{p}^{1\overline{1}\overline{4}}
+\hat{p}^{2\overline{2}\overline{4}}+\hat{p}^{3\overline{3}\overline{4}})\vert 0\rangle.
\label{kanghz72}
\eeq
\noindent
This state is labelled by the set of parameters
\beq
\alpha,A,B,\beta,D,E,F)=(\eta,X,Y,\xi,Z,V,U)=(1,I,0,\xi,I,0,U),
\label{set2}
\eeq
\noindent
i.e. we have taken the negative of the matrix $X$.
As a result of this our matrix $\mathcal{B}$ takes the following form
\beq
\mathcal{B}_{IJ}(\Phi_+)=\begin{pmatrix}-I&-\frac{1}{2}(\xi I-U)&-\frac{1}{2}u\\-\frac{1}{2}(\xi I+U)&I&0\\
-\frac{1}{2}u^T&0&1\end{pmatrix}.
\label{calbematrixspec}
\eeq
\noindent
A diagonalization of $\mathcal{B}_{IJ}(\Phi_+)$ yields the eigenvalues
\beq
\lambda_{1,2,3}=-\sqrt{1+\frac{1}{4}Q^2},\qquad
\lambda_{4,5,6}=\sqrt{1+\frac{1}{4}Q^2},\qquad \lambda_7=1.
\label{ujsvektorok}
\eeq
\noindent
Then according to Eq.(\ref{kobinv}) we have $\mathcal{J}(\Phi_+)=-\left(1+\frac{1}{4}Q^2\right)$.
Hence for $\xi$ and $U$ real we cannot leave the dense orbit.
However, for complex parameters if the constraint $Q^2=-4$ is satisfied then
the rank of $\mathcal{B}$ is again changing from seven to four.
In order to check this just take $(\xi,u^{\overline{1}},u^{\overline{2}},u^{\overline{3}})=(2i,0,0,0)$.
In this case the matrix of Eq.(\ref{calbematrixspec}) has a block diaginal structure containing a $6\times 6$ block of the form $-(\sigma_3+i\sigma_1)\otimes I$ where $\sigma_{1,3}$ are the usual Pauli matrices.
However $\sigma_3+i\sigma_1$ is  of rank one, hence the $6\times 6$ block is of rank three. Hence the matrix
$\mathcal{B}(\Phi_+)$ is of rank four.
Then if the constraint $Q^2=-4$ holds the state $\vert\Phi_+\rangle$ will again belong to class IX.

The upshot of these considerations is as follows.
Taking the perturbation parameters $\xi,u^{\overline{1}},u^{\overline{2}},u^{\overline{3}}$ to be real reveals an interesting effect.
If we change the sign of the $X$  parameters of the doubles (see Eqs.(\ref{set1}) and (\ref{set2}) )  then
the perturbation due to triples
cannot induce a transition to a different SLOCC class.
Hence in this special case the entanglement (which is according to Table 1. is of type X. ) encoded into the parameters 
of the doubles is protected from the perturbating effect of the triples.
On the other hand if no sign change occurs then if the perturbation parameters are constrained to lie on a three dimensional sphere of radius two, then the entanglement type is transformed to class IX.
Note that the different behavior of the states $\vert\Psi_-\rangle$ and $\vert\Psi_+\rangle$ under perturbations with {\it real} parameters is related to the fact that these states are equivalent under complex SLOCC transformations but {\it inequivalent} under real ones\cite{Hitchin}.

\section{The role of doubles in characterizing entanglement}

Let us now address the entanglement aspects of the CC method in a somewhat more general setting. 
For an arbitrary number of fermions ($n$) and single particle states ($N$)
one can use the coupled cluster expansion of Eq.(\ref{vesszosalak}).  
As we know from this expansion
the contribution coming from singles can be removed since this merely amounts to a SLOCC transformation.
Hence for the characterization of entanglement we are left with the the terms featuring the exponentials of cluster operators $\hat{T}_2,\hat{T}_3,\dots \hat{T}_n$. In this section we would like to put forward the proposal to characterize fermionic entanglement by keeping merely one of such exponentials namely the contribution coming from doubles: $\hat{T}_2$.
Explicitly let us consider the approximation to $\vert \Psi\rangle$ of  Eq.(\ref{vesszosalak}) 
in the form
\beq
\vert\Phi\rangle=e^{\hat{T}_2}\vert\Psi_0\rangle,\qquad  \hat{T}_2=\frac{1}{4}{T_{ab}}^{ij}\hat{p}^a\hat{n}_i\hat{p}^b\hat{n}_j
\label{csonka}
\eeq
\noindent
where $i,j=1,2,\dots n$ and $a,b=1,2,\dots N-n$.
Naively on expects that for sufficiently small $n$ and $N$ the parametrization based on $\hat{T}_2$ is capable of recovering all the SLOCC classes.
At the same time our  parametrization of fermionic entangled states in terms of parameters of $\hat{T}_2$ is reducing the number of entanglement parameters from ${N\choose n}$ to ${n\choose 2}{N-n\choose 2}$. However, the price we have to pay
is that
 for larger values of $n$ and $N$  
the description of entanglement in terms of doubles can only produce a coarse graining which is generally insensitive to the fine structure of SLOCC classes. At the same time we must bear in mind that though we are losing information on the fine deatails of entanglement but at the same time this approach could be  appealing from the physical point of view.
Indeed, entanglement is a resource and it is the physical problem at hand that defines the criteria under which we should classify this resource.
Since most of the interaction terms used in solid state, molecular and atomic physics are based on interaction terms with similar structure to that of $\hat{T}_2$ our proposal could be a valuable tool.
Of course in order to check the viability of our approach investigations with realistic systems should be considered.
Such investigations we would like to perform in future works.
For the time being in the following we would merely like to offer some solid pieces of mathematical evidence in favour of our proposal.

Let us see first how the usual SLOCC classes are reproduced for our cases of $n=3$, $N=6,7$ in terms of doubles.

In the $N=6$ case according to Eq.(\ref{sixXY}) keeping merely the contribution from the doubles amounts to
using the $3\times 3$ matrix $X$ for entanglement characterization.
According to our discussion of SLOCC classes at the end of Section 2.3 we see that in terms of the matrix $X$
separable states are the trivial ones with $X=X^{\sharp}={\rm Det}X=0$.  
For biseparable states we have $X\neq 0$ but $X^{\sharp}={\rm Det}X=0$. For the classes containing genuine entanglement the W-class has $X\neq 0$ and $X^{\sharp}\neq 0$ but ${\rm Det}X=0$, on the other hand for the GHZ-class none of these three quantities are zero.  
This is in accord with Eq.(\ref{kvartik2}) showing that for the GHZ class $\mathcal{D}(\Phi)={\rm Det}X\neq 0$.
Hence in this special case we see that a parametrization in terms of doubles intersects all of the usual SLOCC classes.

In the $N=7$ case describing entanglement in terms of doubles means a characterization 
in terms of only the two $3\times 3$ matrices 
\beq
{X^a}_i=\frac{1}{4}\varepsilon^{abc}\varepsilon_{ijk}{T_{bc}}^{jk},\qquad
Z_{ka}=\frac{1}{2}\varepsilon_{ijk}{T_{a4}}^{ij}.
\eeq
\noindent
In this case $\mathcal{J}(\Phi)=-{\rm Det}G$ where G is the symmetric part of the matrix $ZX$.
Hence we are in class X. of Table 1. provided this invariant is nonzero. Clearly the classes II.-V. are the ones familiar from the $N=6$ case. In this case $Z=0$, and the constraints are just the ones of the previous paragraph.
In order to identify the constraints in terms of $X$ and $Z$ for the remaining classes of Table 1. one has to look at the ranks of the covariants. In particular for the calculation of the rank of the covariant 
$N_{IJ}$ we have to use the expressions as displayed in the Appendix where now $\xi=0$ and $U=0$.
Notice however, that in order to separate classes VI. and VII. further covariants are needed. For the structure of these covariants see Ref.\cite{SarLev2}.

As a less trivial example let us consider some aspects of entanglement characterization in terms of doubles for the
$n=4$, $N=8$ case.
In this case we have to use the cluster operator $\hat{T}_2$ of Eq.(\ref{csonka}) with $i,j=1,2,3,4$ and
$a,b=\overline{1},\overline{2},\overline{3},\overline{4}$.
In order to calculate $\vert\Phi\rangle$ of Eq.(\ref{csonka}) we calculate
\beq
\vert\Phi\rangle=(1+\hat{T}_2+\frac{1}{2}\hat{T}_2^2)\hat{p}^{1234}\vert 0\rangle
\eeq
\noindent
with
\beq
\hat{T}_2\vert\Psi_0\rangle=\frac{1}{2}({T_{ab}}^{12}\hat{p}^{34}+{T_{ab}}^{34}\hat{p}^{12}-
{T_{ab}}^{13}\hat{p}^{24}-{T_{ab}}^{24}\hat{p}^{13}+
{T_{ab}}^{14}\hat{p}^{23}+{T_{ab}}^{23}\hat{p}^{14}
)\hat{p}^{ab}\vert 0\rangle
\eeq
\noindent
\beq
\frac{1}{2}\hat{T}_2^2\vert\Psi_0\rangle=\frac{1}{4}({T_{ab}}^{12}{T_{cd}}^{34}-
{T_{ab}}^{13}{T_{cd}}^{24}+{T_{ab}}^{14}{T_{cd}}^{23})\varepsilon^{abcd}\hat{p}^{\overline{1234}}\vert 0\rangle.
\eeq
\noindent

Since the structure of the entangelment classes in this case is quite involved\cite{Chen1} we are content with showing that the orbit which is closed under the SLOCC subgroup $SL(8,\mathbb{C})$ (this is a "GHZ-like" class) can be recovered by this parametrization in terms of doubles. Since almost all orbits are closed with respect to this group action\cite{Chen1} hence we can conclude that for all practical purposes for obtaining a coarse grained picture of entanglement this parametrization will do.(It would be interesting to check whether we are still capable of discriminating all the fine structure of SLOCC orbits, as we were in the $n=3, N=3$ case, a problem we are not addressing here.)

In order to show this the only thing we have to notice is that our parametrization with doubles obviously intersects the seven dimensional subspace spanned by the vectors obtained by acting with the following set of operators on the vacuum.\beq
\hat{P}_1=\hat{p}^{1234}+\hat{p}^{\overline{1234}},
\quad
\hat{P}_2=\hat{p}^{13\overline{13}}+\hat{p}^{24\overline{24}},\quad
\hat{P}_3=\hat{p}^{12\overline{12}}+\hat{p}^{34\overline{34}}
\eeq
\noindent
\beq
\hat{P}_4=\hat{p}^{13\overline{24}}+\hat{p}^{{24}\overline{13}},
\quad
\hat{P}_5=\hat{p}^{14\overline{14}}+\hat{p}^{{23}\overline{23}},
\quad
\hat{P}_6=-\hat{p}^{14\overline{23}}-\hat{p}^{{23}\overline{14}},
\quad
\hat{P}_7=-\hat{p}^{12\overline{34}}-\hat{p}^{{34}\overline{12}}.
\eeq
\noindent
According to Ref.\cite{Chen1} the closed orbits are precisely those that meet this subspace 
generated by the vectors above.
Indeed, choosing
\beq
{T_{\overline{12}}}^{12}={T_{\overline{34}}}^{34}=a,\quad
{T_{\overline{12}}}^{34}={T_{\overline{34}}}^{12}=b,\quad
{T_{\overline{13}}}^{13}={T_{\overline{24}}}^{24}=c,\quad
{T_{\overline{24}}}^{13}={T_{\overline{13}}}^{24}=d,
\eeq
\noindent
\beq
{T_{\overline{14}}}^{14}={T_{\overline{23}}}^{23}=e,\quad
{T_{\overline{23}}}^{14}={T_{\overline{14}}}^{23}=f,\quad a^2+b^2-c^2-d^2+e^2+f^2=1
\eeq
\noindent
shows that $\vert\Phi\rangle$ will be a state in the subspace spanned by $\hat{P}_I\vert 0\rangle, I=1,2,\dots 7$.

\section{Conclusions}

In this paper by studying simple fermionic systems 
we established a connection between  the Coupled Cluster (CC) Method
and quantum entanglement.
In the CC method for $n$-fermions with $N$ modes one starts with a single Slater determinant comprising a special set of $n$  single particle states, called the {\it occupied ones}.
Then to this distinguished state one applies a commuting set of cluster operators of the form $e^{\hat{T}_i}, i=1,2,\dots n$.
We have initiated a study of the problem how these cluster operators are encoding the SLOCC entanglement properties of the system via complex expansion coefficients of operator combinations called singles, doubles, triples etc. These operators are describing transitions from the occupied modes to the non-occupied ones.
Our starting point was the simple observation that the operator $e^{\hat{T}_1}$ featuring singles is a special fermionic SLOCC transformation.
Indeed, the transformation $\vert\Psi\rangle\mapsto e^{\hat{T}_1}\vert\Psi\rangle$
can be rewritten in the form $\vert\Psi\rangle\mapsto S\otimes S\otimes\cdots \otimes S\vert\Psi\rangle$
meaning
\beq
\Psi_{\mu_1\mu_2\dots\mu_n}\mapsto {S_{\mu_1}}^{\nu_1}
{S_{\mu_2}}^{\nu_2}\cdots {S_{\mu_n}}^{\nu_n}
\Psi_{\nu_1\nu_2\dots\nu_n},\qquad S\in
GL(N,\mathbb{C}) \label{sixtrafon}
 \eeq
 \noindent
with $S$ an $N\times N$ matrix having the (\ref{trianglematr}) structure.
As a result of this in order to study SLOCC entanglement we merely have to look at the structure of the cluster operators
$\hat{T}_2,\hat{T}_3,\dots \hat{T}_n$.

In order to make some progress in clarifying how multipartite entanglement manifests itself in these operators we conducted a study on the simplest nontrivial cases of $n=3$ and $N=6,7$. In these cases the structure of the SLOCC classes and the structure of the invariants and covariants is explicitely known. 
Surprisingly in the special case of six modes the CC expansion is tailor made
to characterize the usual SLOCC entanglement classes. It means that
the notion of a SLOCC transformation shows up quite naturally as a
one relating the CC and CI expansions
, and going from the CI
expansion to the CC one is equivalent to obtaining a
canonical form for the state where the structure of
the entanglement classes is transparent.
Using the CC parametrization of states (Eqs.(\ref{sixC})-(\ref{sixXY}) and Eqs.(\ref{ujCI1})-(\ref{matrixvector})) in the six and seven mode cases we have given
simple expressions for the unique SLOCC invariants $\mathcal{D}$ and  $\mathcal{J}$ (see Eqs.(\ref{kvartik2}) and (\ref{klassz})) serving as natural measures of multipartite entanglement\cite{LevVran,SarLevCKW,SarLev2}.
In the six mode case we have given a full characteriation of the SLOCC entanglment classes in terms of the cluster operators corresponding to doubles and triples.
In the seven mode case we have considered a perturbation problem featuring a state from the
unique SLOCC class characterized by $\mathcal{J}\neq 0$.
For this state
with entanglement generated by doubles
we investigated the phenomenon of changing the entanglement type due to the perturbing effect of triples.
For the representative of the dense orbit (class X.) we have chosen a "GHZ-like" state generated by doubles.
We then used the four complex parameters characterizing the triples as perturbations. We have shown that when these parameters define a deformed conifold with a deformation parameter $Q=2$ then a transition to a new entanglement class (class IX.) is possible.
We have also demonstrated that there are states with real amplitudes such that their entanglement encoded into configurations of clusters of doubles is protected from errors generated by triples.

The results of this paper were primary aimed at drawing the readers attention to the existence of the connection between multipartite fermionic entanglement and the structure of coupled cluster operators.
A natural question to be asked is how our results illustrated merely in the simplest nontrivial multipartite (i.e. tripartite) cases generalize to more realistic ones which are containing genuine multipartite systems with an arbitrary number of modes.
A promising idea in this respect seems to consider entanglement characterization via doubles meaning by the entanglement parameters showing up in the cluster operator $\hat{T}_2$.
Dealing with merely doubles is reducing the entanglement parameters that we have to account for, but at the same time we are loosing some of the finer details provided by the usual SLOCC classification.
However, entanglement is a resource and it is the physical problem at hand that defines the criteria under which we should classify this resource.
Since interactions based on exchange processes involving pairs of modes (particles) are common in physics, an appreciation of this idea could provide a basis for further elaborations of our proposal. 

In order to fully explore the connection found here between entnaglement and the CC-method there are many possible avenues to follow.
Obviously the most important of such vistas would be a demonstration how the entanglement encoded in  the cluster operators can be used for studying the properties of realistic (e.g. three electron) systems.
Such ideas we are intending to explore in future work.
Another mainly theoretical line of development
would be to invoke further results from the theory of spinors\cite{Cartan,Chevalley,Trautman,Trautmanfia}.
In particular in order to investigate entanglement patterns one has to use the natural invariants defined for spinors\cite{Chevalley,Trautman,Trautmanfia,SarLev3,LevHolw}, and possibly other notions such as the nullity of a spinor\cite{Trautmanfia}. Hopefully these structures will provide a further insight into the coarse grained structure of entanglement types based on the CC method.

\section{Appendix}

Here for the convenience of the reader
for the seven single particle case
we give the explicit form of the covariants $N_{IJ}$ and $L^{IJ}$ in terms of the relevant
coupled cluster parameters $(\xi, X,Z,U)$. In the expressions below $X$ and $Z$ and $U$ are complex $3\times 3$ matrices where $U$ is antisymmetric hence it can also be parametrized as $U_{ab}=\varepsilon_{abc}u^c$ with
$u^c$ a three component vector.
Recall the index structures  $Z_{ia}$, ${X^a}_i$ , $(Z^{\sharp})^{ai}$ and ${(X^{\sharp})^i}_a$ which are being in accord with expressions like $Z_{ia}(Z^{\sharp})^{aj}=({\rm Det Z}){\delta_i}^j$ etc.

\beq
N_{ij}=3[ZX+(ZX)^T]_{ij},\qquad N_{ab}=-3[Z^TX^{\sharp}+(Z^TX^{\sharp})^T]_{ab},\qquad N_{77}=-6{\rm Det}Z
\label{N1}
\eeq
\noindent
\beq
N_{ai}=3(UX+\xi Z^T)_{ai},\qquad N_{i7}=3(Zu)_i,\qquad N_{a7}= -3\varepsilon_{abc}(XZ^{T\sharp})^{bc}.
\label{N2}
\eeq
\noindent
\beq
L^{ij}=-6[(ZX)^{\sharp}+(ZX)^{T\sharp}]-3\varepsilon^{ikl}\varepsilon^{jmn}[(ZX)_{km}(ZX)^T_{ln}+
(ZX)^T_{km}(ZX)_{ln}]
\label{L1}
\eeq
\noindent
\beq
L^{ab}=12[(XZ^{T\sharp})+(XZ^{T\sharp})^T]^{ab}+6u^au^b,\qquad L^{77}=6[\xi^2+4{\rm Det}X].
\label{L2}
\eeq
\noindent
\beq
L^{ai}=12\varepsilon^{ijk}{X^a}_j(Zu)_k+6\varepsilon^{ijk}u^a(ZX)_{jk}-12\xi(Z^{\sharp})^{ai}
\label{L3}
\eeq
\noindent
\beq
L^{i7}=6\xi\varepsilon^{ijk}(ZX)_{jk}-6\varepsilon^{ijk}U_{ab}{X^a}_j{X^b}_k,\qquad
L^{a7}=-6\xi u^a+12\varepsilon^{abc}(Z^TX^{\sharp})_{bc}.
\label{L4}
\eeq

\end{document}